# On the complexity of deciding whether the distinguishing chromatic number of a graph is at most two

Elaine M. Eschen [*]   Chính T. Hoàng [†]   R. Sritharan [‡]

Lorna Stewart [§]

October 29, 2018

## Abstract

In an article [3] published recently in this journal, it was shown that when $k \geq 3$, the problem of deciding whether the distinguishing chromatic number of a graph is at most $k$ is NP-hard. We consider the problem when $k = 2$. In regards to the issue of solvability in polynomial time, we show that the problem is at least as hard as graph automorphism but no harder than graph isomorphism.

## 1 Introduction

We consider simple undirected graphs. A *nontrivial* automorphism of a graph is an automorphism that is not the identity mapping. We use the abbreviation NTA for nontrivial automorphism. A graph that has no nontrivial automorphism is said to be *asymmetric*. A vertex $k$-coloring of graph $G = (V, E)$ is a mapping $V \to \{1, 2, \cdots, k\}$. A vertex $k$-coloring of graph $G$ is *proper* if no two adjacent vertices of $G$ receive the same color. $\chi(G)$

<>[*]elaine.eschen@mail.wvu.edu, Lane Department of Computer Science and Electrical Engineering, West Virginia University.
   [†]choang@wlu.ca, Department of Physics and Computer Science, Wilfrid Laurier University, Waterloo, Canada. Acknowledges support from NSERC of Canada.
   [‡]srithara@notes.udayton.edu, Computer Science Department, The University of Dayton, Dayton, OH 45469. Acknowledges support from the National Security Agency, USA.
   [§]stewart@cs.ualberta.ca, Department of Computing Science, University of Alberta, Edmonton, Alberta, Canada T6G 2E8. Acknowledges support from NSERC of Canada.

is the chromatic number of graph $G$, namely, the smallest positive integer $k$ such that $G$ admits a proper vertex $k$-coloring. We use $\leq_m$ and $\equiv_m$ to denote polynomial-time many-one reducibiliy and equivalence, respectively, and $\leq_T$ for polynomial-time Turing reducibility.

A vertex $k$-coloring of graph $G$ is *distinguishing* if the only automorphism of $G$ that preserves the coloring is the identity automorphism. The *distinguishing number of graph $G$*, denoted $D(G)$, is the smallest positive integer $k$ such that $G$ admits a $k$-coloring (not necessarily proper) that is distinguishing. Similarly, the *distinguishing chromatic number of graph $G$*, denoted $\chi_D(G)$, is the smallest positive integer $k$ such that $G$ admits a *proper $k$-coloring* that is distinguishing. The concept of distinguishing number of a graph was introduced by Albertson and Collins in [1]. Later, Collins and Trenk [4] introduced the notion of distinguishing chromatic numbers of graphs.

The computational complexities of the problems of computing $D(G)$ and $\chi_D(G)$ have been investigated in the recent past. It was shown in [7] that given a graph $G$ and integer $k$, deciding whether $D(G) \leq k$ belongs to AM, the set of languages for which there exist Arthur and Merlin games. In a more recent paper [2] it has been shown that given a planar graph $G$ and an integer $k$, whether $D(G) \leq k$ can be decided in polynomial time. Cheng [3] has shown that given an interval graph $G$ and an integer $k$, whether $\chi_D(G) \leq k$ can be decided in polynomial time. In contrast to this, Cheng [3] also established that given an arbitrary graph $G$ and an integer $k$, where $k \geq 3$, deciding whether $\chi_D(G) \leq k$ is NP-hard. Further, the problem remains NP-hard when $k = 3$ and the input graph is planar with maximum degree at most five [3]. In regards to the problem of deciding whether $\chi_D(G) \leq 2$, given a graph $G$, Cheng remarked in [3] that "it will be interesting to consider what the corresponding results are" for deciding whether $\chi_D(G) \leq 2$.

We show that given a *connected* graph $G$, deciding whether $\chi_D(G) \leq 2$ is polynomial-time Turing equivalent to the problem of deciding whether a given graph $H$ has a NTA. Thus, given an *arbitrary* graph $G$, deciding whether $\chi_D(G) \leq 2$ is at least as hard as deciding whether a graph $H$ has any NTA. We then show that given an arbitrary graph $G$, the problem of deciding whether $\chi_D(G) \leq 2$ is no harder than deciding whether given graphs $G_1$ and $G_2$ are isomorphic to each other.

Next, we introduce the definitions of some needed problems. Then, we present our main results. Finally, we conclude with some discussion.

## 2  Graph automorphism and graph isomorphism

Consider the following decision problems each of which is known to be in NP, but neither of which is known to be in P or NP-complete. Graph isomorphism has long been considered a candidate to be in NP but neither in P nor NP-complete (such problems are known to exist if P $\neq$ NP [6]).

> GRAPH AUTOMORPHISM (**GA**)
> Instance: Graph $G$.
> Question: Does $G$ have a nontrivial automorphism?
>
> GRAPH ISOMPORPHISM (**GI**)
> Instance: Graphs $G_1$ and $G_2$.
> Question: Is $G_1 \cong G_2$?

It is known that **GA** $\leq_m$ **GI** [5]; however, as stated in [5], "**GI** does not seem to be reducible to **GA**". Thus, it is possible that **GA** is easier to compute than **GI**.

## 3  Results

It can be observed based on the definitions that $\chi(G) \leq \chi_D(G)$ and that $D(G) \leq \chi_D(G)$. If $G$ is asymmetric, then $\chi_D(G) = \chi(G)$. Clearly, $D(G) = 1$ if and only if $G$ is asymmetric. Therefore, given graph $G$, deciding whether $D(G) = 1$ is polynomial-time equivalent to **GA**. In contrast, given graph $G$, deciding whether $\chi_D(G) = 1$ is trivial; $G = K_1$ is the only graph with $\chi_D(G) = 1$. When $\chi_D(G) = 2$, $G$ is necessarily bipartite. In the remainder of the paper, we use 2-coloring to refer to a proper 2-coloring.

Our focus is on the following problem:

> DISTINGUISHING 2-COLORABILITY (**D2C**)
> Instance: Graph $G$.
> Question: Is $\chi_D(G) \leq 2$?

We first consider the problem **D2C** restricted to connected graphs. Note that a connected bipartite graph $G$ has a unique (up to renaming the colors) 2-coloring and, therefore, either every 2-coloring of $G$ is distinguishing or none of them is. Consequently, **D2C** for connected graphs is polynomial-time many-one equivalent to the problem: Given a graph $G$ and a 2-coloring $c$ of $G$, is $c$ a distinguishing coloring? Thus, since a given coloring $c$ of graph $G$ is *not* distinguishing if and only if there is a NTA of $G$ that preserves $c$, the complement of **D2C** restricted to connected graphs seems closely related to **GA**. The next theorem shows that those two problems are in fact polynomial-time many-one equivalent.

In the remainder of the paper, we refer to the *complement* of **D2C** restricted to *connected* graphs as **CC**:

> COMPLEMENT OF **D2C** ON CONNECTED GRAPHS (**CC**)
> Instance: *Connected* graph $G$.
> Question: Is $\chi_D(G) \nleq 2$?

**Theorem 1** *Problems* **CC** *and* **GA** *are polynomial-time many-one equivalent.*

**Proof.** First, we show that **GA** $\leq_m$ **CC**.

Since a graph has a NTA if and only if its complement has a NTA, and the complement of a disconnected graph is connected, we may assume that the given instance $G = (V, E)$ of **GA** is connected.

If $G = K_1$, $G$ is a NO instance of **GA** and we can easily construct a NO instance $G'$ of **CC**. Otherwise, let $G' = (V' = V \cup E, E')$ be the graph obtained from $G$ by subdiving each edge of $G$ once. We note that for an edge $xy$ of $G$, we use $xy$ to refer to the edge of $G$ as well as the vertex of $G'$ that subdivides the edge $xy$ of $G$. Clearly, $G'$ is connected and bipartite, and every vertex in $E$ has degree 2. In order to complete the reduction from **GA** to **CC**, we prove that $G$ has a NTA if and only if $\chi_D(G') \nleq 2$.

If all the vertices of $V'$ have degree 2 then $G$ is a chordless cycle of length $\geq 3$ and therefore has a NTA. In this case, for every 2-coloring of $G'$, the vertices in $V$ are mapped to one color, the vertices in $E$ are mapped to the other color, and also there is a NTA of $G'$ that preserves the coloring. Therefore, $G$ has a NTA and $\chi_D(G') \nleq 2$.

In the remaining case, one color class of $G'$ consists entirely of degree 2 vertices (vertices in $E$) and the other color class (vertices in $V$) contains a vertex of degree $\neq 2$. Thus, every NTA of $G'$ must map $V$ to $V$ and $E$ to $E$.

First, we show that if $f : V \mapsto V$ is a NTA of $G$, then for every 2-coloring of $G'$ there exists a NTA $f' : V' \mapsto V'$ that preserves the coloring of $G'$ (and hence $\chi_D(G') \nleq 2$). Note that as $G'$ is connected, it is enough to consider a particular 2-coloring of $G'$.

Suppose $f : V \mapsto V$ is a NTA of $G$. Define $f' : V' \mapsto V'$ where

$$f'(x) = \begin{cases} f(x) & \text{if } x \in V \\ f(u)f(v) & \text{if } x = uv \in E \end{cases}$$

We now show that $f'$ is an NTA that preserves every 2-coloring of $G'$. Since $f$ is an automorphism, $f'$ is a bijection. To see that $f'$ is an automorphism, observe that:

$$\begin{aligned} uv \in E' &\Leftrightarrow u \in V \text{ and } v \in E \text{ and } v = uw \text{ for some } w \in V \text{ (or vice versa)} \\ &\Leftrightarrow f'(u) = f(u) \in V \text{ and } f'(v) = f(u)f(w) \\ &\Leftrightarrow f'(u)f'(v) \in E' \text{ since } f'(v) \text{ corresponds to an element of } E \\ &\phantom{\Leftrightarrow} \text{ that is incident with } f'(u) \text{ in } G. \end{aligned}$$

Since $G'$ is connected, it has a unique 2-coloring, and that 2-coloring is preserved by $f'$ since $f'$ maps $V$ to $V$ and $E$ to $E$. Finally, as $f$ is a NTA of $G$, $f'$ is a NTA of $G'$.

Next, we show that if $\chi_D(G') \not\leq 2$, then $G$ has a NTA. Suppose $\chi_D(G') \not\leq 2$. Let $c$ be the unique 2-coloring of $G'$ and let $f' : V' \mapsto V'$ be a NTA of $G'$ that preserves $c$. Define $f : V \mapsto V$ such that $f(x) = f'(x)$ for all $x \in V$. Since $f'$ preserves $c$, it maps $V$ to $V$ and $E$ to $E$. As $f'$ is a NTA of $G'$, the $V$ to $V$ mapping of $f$ is a NTA of $G$.

This completes the proof of **GA** $\leq_m$ **CC**.

We now prove the reduction in the other direction, that is, **CC** $\leq_m$ **GA**. Let $G = (X, Y, E)$ be a connected bipartite graph that is not $K_1$ or $K_2$. ($K_1$ and $K_2$ are NO instances of **CC** and any connected non-bipartite graph $G$ is an YES instance of **CC**. In these cases, we can construct $G'$ accordingly.) Note that $G$ has a unique 2-coloring with color classes $X$ and $Y$. Define

$$G' = \begin{cases} G = (X, Y, E) & \text{if } |X| \neq |Y| \\ (X', Y', E') & \text{otherwise} \end{cases}$$

where $a, b, c \notin X \cup Y$ and

$$\begin{aligned} X' &= X \cup \{b\} \\ Y' &= Y \cup \{a, c\} \\ E' &= E \cup \{ax \mid x \in X\} \cup \{ab, bc\} \end{aligned}$$

We prove that $\chi_D(G) \not\leq 2$ if and only if $G'$ has a NTA.

Suppose $\chi_D(G) \not\leq 2$. Then there exists a NTA $f$ of $G$ that preserves the unique 2-coloring of $G$. In the case that $G' = G$, $f$ is also a NTA of $G'$. In the case that $G' \neq G$ define the mapping $f' : X' \cup Y' \mapsto X' \cup Y'$ where

$$f'(x) = \begin{cases} f(x) & \text{if } x \in X \cup Y \\ x & \text{if } x \in \{a, b, c\}. \end{cases}$$

It is easily seen that $f'$ is a NTA of $G'$.

Now suppose $f$ is a NTA of $G' = (X', Y', E')$. Since $|X'| \neq |Y'|$ and $G'$ is connected, $f$ preserves the unique 2-coloring of $G'$. Further, $f(a) = a$, $f(b) = b$, and $f(c) = c$ by the vertex degrees, the connectedness of $G$, and the fact that $G \not\cong K_2$. Thus, $f$ maps $X$ to $X$ and $Y$ to $Y$ and therefore $f$ restricted to $G$ is a NTA of $G$ that preserves the unique 2-coloring of $G$. Therefore, $\chi_D(G) \not\leq 2$ and the proof of the theorem is complete. ■

The following proposition allows us to analyze the complexity of problem **D2C** for graphs that are not necessarily connected. We again use the fact that a connected bipartite graph has a unique (up to renaming the colors) 2-coloring.

**Proposition 1** *Let $G$ be a graph. $\chi_D(G) \leq 2$ if and only if*

- *$G$ is bipartite and*

- *for every component $C$ of $G$:*

    - *$\chi_D(C) \leq 2$,*
    - *$C$ is isomorphic to at most one other component of $G$, and*
    - *if $C$ is isomorphic to some other component of $G$ then $C$ is asymmetric.*

**Proof.** The proposition clearly holds when $G = K_1$. Therefore, we now assume $G$ has at least two vertices.

$\Rightarrow$ Suppose that $\chi_D(G) \leq 2$. Then, $G$ is bipartite by an earlier observation. By the definition of $\chi_D$, there is a 2-coloring $c$ of $G$ that is distinguishing.

Let $C$ be a component of $G$. It is clear that $c$ restricted to $C$ is a distinguishing coloring or else we contradict the choice of $c$.

Suppose that $C_1 = (X_1, Y_1, E_1)$, $C_2 = (X_2, Y_2, E_2)$, and $C_3 = (X_3, Y_3, E_3)$ are three distinct isomorphic components of $G$ and that there are isomorphisms mapping $X_1$ to $X_2$ and $X_2$ to $X_3$. No matter how the vertices of $C_1$, $C_2$, and $C_3$ are 2-colored, two of $X_1, X_2, X_3$ will be in the same color class and therefore for every 2-coloring of $G$, there is an NTA that preserves the coloring (specifically, an automorphism that maps the two $X_i$'s that are in the same color class to one another), contradicting that $\chi_D(G) \leq 2$. Therefore, each component can be isomorphic to at most one other component.

Suppose that $C = (X_C, Y_C, E_C)$ is isomorphic to another component $C' = (X_{C'}, Y_{C'}, E_{C'})$ and that some isomorphism $f$ maps $X_C$ to $X_{C'}$. If $C$ is not asymmetric, then it has a NTA $g$, and every NTA of $C$ maps $X_C$ to $Y_C$ and vice versa, or else we contradict that $c$ is distinguishing. But, now there are two isomorphisms from $C$ to $C'$, namely, $f$ and $g \circ f$, one of which preserves $c$, a contradiction.

$\Leftarrow$ Let bipartite graph $G$, with components $C_1, C_2, \ldots, C_k$, satisfy the conditions. Suppose $\chi_D(G) \nleq 2$. Then, for every 2-coloring of $G$, there is a NTA that preserves the coloring. Let $c$ be a 2-coloring of $G$ in which isomorphic pairs of components are colored such that if there is an isomorphism mapping $X_1$ to $X_2$ then $X_1$ and $X_2$ have opposite colors in $c$. Now, every NTA swaps colors within single components and/or swaps colors in pairs of components but, in any case, $c$ is not preserved, a contradiction. ∎

**Corollary 1** **D2C** $\leq_T$ **GI**

**Proof.**

By Propositon 1, an algorithm for **D2C** can be constructed from algorithms for **CC**, **GI**, and **GA**. Since **CC** $\leq_m$ **GA** (Theorem 1) and **GA** $\leq_m$ **GI**, the result follows. ∎

## 4 Discussion

Combining Theorem 1, Corollary 1, and the observation that **CC** $\leq_T$ **D2C**, we have **GA** $\equiv_m$ **CC** $\leq_T$ **D2C** $\leq_T$ **GI**. That is, **D2C** is at least as hard as **GA** and no harder than **GI**, in terms of Turing reductions.

Our results imply that **CC** $\in$ NP and **D2C** $\in$ co-NP. In addition, a direct consequence of Corollary 1 is that for a graph $G$ belonging to a class $\mathcal{C}$ such that the isomorphism problem can be solved in polynomial time for $\mathcal{C}$, deciding whether $\chi_D(G) \leq 2$ can be done in polynomial time.

A question that arises from Theorem 1 and Corollary 1 is: is problem **D2C** polynomial-time equivalent to **GA** or to **GI**, or does its complexity lie in between those of problems **GA** and **GI**?